# Special relativity with clock synchronization

Bernhard Rothenstein[1], Stefan Popescu[2] and George J. Spix[3]

1) Politehnica University of Timisoara, Physics Department, Timisoara, Romania, bernhard_rothenstein@yahoo.com
2) Siemens AG, Erlangen, Germany, stefan.popescu@siemens.com
3) BSEE Illinois Institute of Technology, USA, gjspix@msn.com

**Abstract.** *We present an introduction to special relativity kinematics stressing the part played by clocks synchronized following a procedure proposed by Einstein.*

### A. Special relativity with clock synchronization
### A.1. Introduction
### A.1.1 The event and its space-time coordinates.

The concept of *event* is of fundamental importance to special relativity theory. We define it as any physical phenomenon that take place at a given point in space at a given time. We define the point where the event takes place by its space coordinates (Cartesian or polar) and the time when it takes place by the reading of a clock located at that point when the event takes place. The physicist who observes nature by conducting experiments is by definition an *observer*. He knows the laws of nature by heart and he is able to operate measuring devices. Einstein's observers work as a team. At each point in space we find such an observer. A first type of observer collects information about the events that take place immediately in front of him. Knowing his location in space using meter sticks and using a clock he characterizes an event by the space coordinates defining his location in space and by a time coordinate that equates the reading of his clock when the event takes place. For more clarity, let say that the clock defined above is the observer's wrist watch. Another type of observer collects information about the space-time coordinates of events that take place remotely at different points in space from the light signals that arrive at his location[1,2] or from the light signal that he sends out and receives back after reflection on a mirror located where the event takes place[3,4].

The observer works in a laboratory. He can be confined to it or he can be an astronomer who contemplates the sky. The observer attaches a *reference frame* K(XOY) to his laboratory that represents its *rest frame* (Figure 1a). We say that the laboratory and the reference frame attached to it are *inertial* if when an object at rest is placed at an arbitrary point, it will continue to stay there. The reference frame enables the observer to assign space coordinates to the points where events take place. The position of point $M$ is defined, in a two space dimensions approach, by its Cartesian ($x,y$) and by its polar ($r,\theta$) space coordinates related by

$$x = r\cos\theta \tag{1.1}$$
$$y = r\sin\theta \tag{1.2}$$
$$r = \sqrt{x^2 + y^2} \tag{1.3}$$

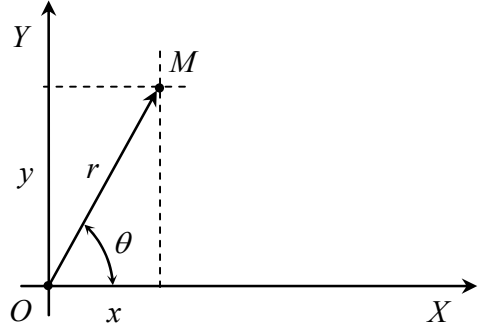

**Figure 1a**. *Defining the position of a point M relative to a reference frame K(XOY) using Cartesian and polar coordinates as well.*

### A.1.2 The relativistic postulates

Consider two inertial reference frames K(XOY) and K'(X'O'Y'). The corresponding axes of the two frames are parallel to each other, the OX(O'X') axes are overlapped and K' moves with constant speed $V$ in the positive direction of the OX(O'X') axes. $C_0(0,0)$ is the wrist watch of an observer $R_0(0,0)$ at rest in K and located at its origin O', whereas $C'_0(0,0)$ is the wrist watch of an observer $R'_0(0,0)$ at rest in K' and located at its origin O'. When the two origins O and O' are located at the same point in space the watches are set to read a zero time. We say that under such conditions the two reference frames are in the *standard arrangement*. We can achieve this by placing a source of light in the immediate vicinity of the stopped clock $C_0$. Clock $C'_0$ stopped as well comes along the OX axis. Both clocks display a zero time. A photoelectric device detects the arrival of clock $C'_0$ at the location of clock $C_0$, closing the electric circuit of a light source that emits light signals perpendicular to the XOY plane starting the two clocks.

Observers of the two frames say $R(x,y)$ and respectively $R'(x',y')$, perform measurements in order to find out the physical properties of a *physical system*. The result of a physical measurement is a *physical quantity* and we express it as a product of a numerical value and a physical unit (where S.I. units are usually preferred). The reference frame relative to which the physical system is in a state of rest represents its rest frame. A *proper physical quantity* represents the result of a measurement performed on a physical system by observers at rest relative to it.

Special relativity theory becomes involved when we try to establish a relationship between the physical quantities, characterizing the same physical system, as measured by observers of K and respectively K'. An equation that mediates these physical quantities is a *transformation equation*. A physical quantity that has the same magnitude when measured from different inertial reference frames represents a *relativistic invariant*. Performing measurements on the same physical object we obtain a variety of physical quantities. An equation that mediates a relationship between them represents a *physical law*. Established in a given reference frame, a law mediates the magnitudes of different physical quantities that characterize the physical properties of the same physical object measured in that frame. If we perform the transformation of each physical quantity present in a physical law in accordance with its transformation equation and the algebraic structure of the law does not change, we say that it is *covariant* with respect to all the transformation equations, which, in this case, constitute a *group*.

The derivation of a transformation equation starts with the statement of *Einstein's postulates* on which special relativity theory is based:

**First postulate**
1) Observation of physical phenomena by more then one inertial observer must result in agreement between the observers as to the nature of reality, or the nature of the universe must not change for inertial observers in relative motion.
2) Every physical theory should look the same mathematically to every inertial observer.
3) No property of the universe will change if the observer is in uniform motion. The laws of the universe are the same regardless of inertial reference frame.
4) A student who has studied physics in the inertial laboratory of a given university will pass the exam at each other laboratory moving uniformly relative to the first one.
5) It is impossible, by experiments performed confined in an inertial reference frame, to decide if the frame is in a state of rest or in a state of uniform motion along a straight line.

6) If the reference frame K' moves with constant velocity $V$ in the positive direction of the OX axis, then K moves with constant velocity $-V$ relative to K' in the negative direction of the same axis.
7) Distances measured perpendicular to the direction of relative motion have the same magnitude in all inertial reference frames in relative motion.[5] The simplest explanation of this consists of the fact that, in the case of the standard arrangement, the relative velocity has no component in the normal direction and so we consider that

$$y = y' \:. \qquad (1.4)$$

The first postulate is an essential extension of Galileo's postulate in which most physicists hardly believe.

**Second postulate**
1) The speed of light in vacuum, commonly denoted as $c$, is the same in all directions and does not depend on the velocity of the object emitting the light.

As a corollary, we argue that when performing an experiment the physicist is confronted with the universe. The universe defends itself, preventing the observers from obtaining results beyond certain limits imposed by the accuracy and capabilities of the available measurement devices, some of these restrictions being actually described by the Einstein's postulates.

Ockham[6] teaches us that if two theories explain the facts equally well, then the simpler theory is the better one. Paraphrasing Ockham, we say that if the same theory explains the same fact equally well, then the shorter explanation is the better one. Our teaching experience convinced us that it is easier to teach those who know nothing about the subject than those with wrong previous representations about it.

For an easier comprehension of this paper it is essential for the reader to admit the Einstein's postulates even if they potentially challenge some instinctual belief developed during previous study of physics. The acceptance of these postulates is supported by many experiments performed in last years as well as by the fact that until today no experiment was possible conducted to contradict these postulates.[7,8]

## A.1.3 The light clock and the other clocks associated with it. Time dilation and length contraction

A *clock* is a physical device that generates a periodic phenomenon of constant period *T*. The best known clock for relativistic experiments is the *light clock*.[9] We present it in its rest frame K' (Figure 1b). The clock consists of two mirrors $M'_1$ and $M'_2$ parallel to the common OX(O'X') axes and located at a distance *d* from each other. A light signal originating from mirror $M'_1$ is reflected back by mirror $M'_2$ and finally returns to the location of $M'_1$. The time interval

$$T' = \frac{2d}{c} \tag{1.5}$$

during which the light signal bounces back and forth between the two mirrors represents the *period* of such a clock.

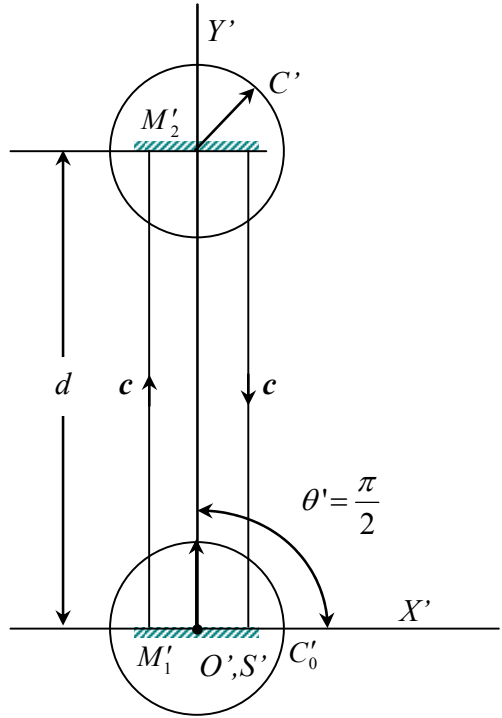

**Figure 1b**. *The light clock in its rest frame K' and the clocks $C'_0(0,0)$ and $C'(0,0)$ associated with it.*

The invariance of the speed of light in vacuum *c* and of distances measured perpendicular to the direction of relative motion makes such a clock very tempting for relativistic speculations. Usually the authors fail to mention that we can attach to the experimental device described above a second clock $C'_0(0,0)$ located in front of mirror $M'_1$ and a third clock $C'(0,d)$ located in front of mirror $M'_2$. The light signals that bounce between the two mirrors could perform the synchronization of the two additional clocks in accordance with the clock synchronization procedure proposed by Einstein.[9] Consider that clock $C'_0(0,0)$ reads $t' = 0$ when a source of light $S'(0,0)$ located in front of mirror $M'_1$ starts to emit light in all directions of the X'O'Y' plane. Previously the clock $C'(0,d)$ is stopped and set to read $t' = \frac{d}{c}$. The light signal that propagates in the positive direction of the O'Y' axis starts the clock $C'(0,d)$ just when arriving at its location. From this very moment the two clocks display the same running time. The reflected light signal returns to clock $C'_0(0,0)$ when it reads $2t' = \frac{2d}{c}$ and at this very moment the clock $C'(0,d)$ reads the same time. Afterwards both clocks will read the same running time.

Special relativity becomes involved when we detect the synchronization of clocks $C'_0(0,0)$ and $C'(0,d)$ from the reference frame K(XOY). The light signals emitted by the source $S'(0,0)$ or by a source $S(0,0)$ at rest in K and located at its origin O, could perform the synchronization of the clocks in the reference frame K(XOY). Let $C_0(0,0)$ be a clock of the reference frame K(XOY) located at its origin O(0,0) and let $C(x = r\cos\theta, y = r\sin\theta)$ be another arbitrary clock of that frame. All clocks read $t = t' = 0$ when the origins O and O' are located at the same point in space.

We watch the synchronization of clocks $C'_0(0,0)$ and $C'(0,d)$ from the reference frame K(XOY). We show the result in Figure 2.

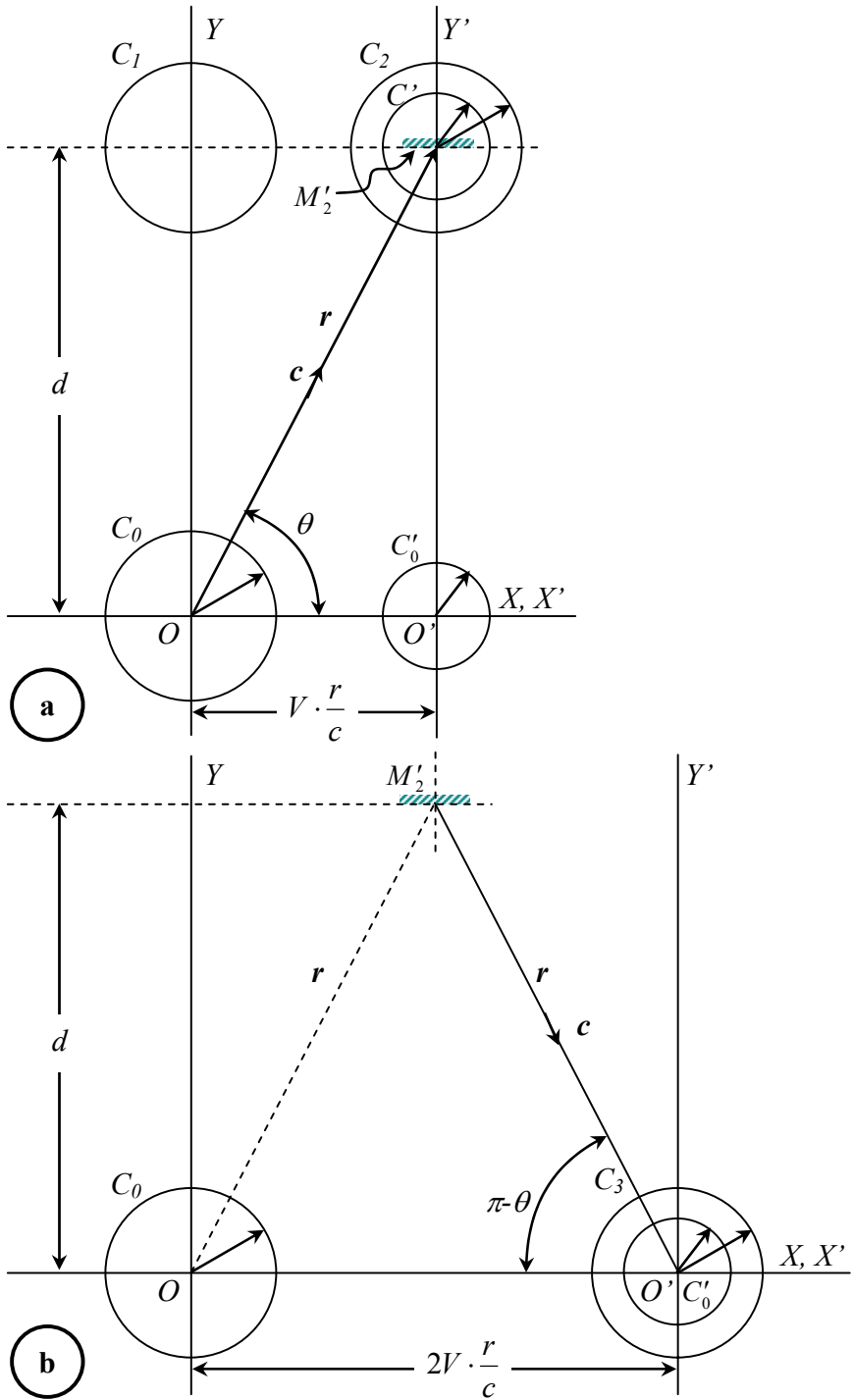

**Figure 2**. *The synchronization of two clocks of the K'(X'O'Y') reference frame as detected from the K(XOY) reference frame.*

Figure 2.a shows the situation when clock $C'(0,d)$, commoving with mirror $M'_2$, arrives in front of a clock $C_2(x = r\cos\theta, y = r\sin\theta)$ of the frame

K(XOY) which is synchronized with $C_0(0,0)$ by a light signal emitted by $S(0,0)$ at $t=0$ along a direction that makes an angle $\theta$ with the positive direction of the common OX(O'X') axes. When clocks $C'(0,d)$ and $C_1(x = r\cos\theta, y = r\sin\theta)$ meet each other, the first reads $t' = \dfrac{d}{c}$ while the second reads $t = \dfrac{r}{c}$. Pythagoras' theorem applied to Figure 2.a leads to

$$r^2 = d^2 + V^2 \frac{r^2}{c^2} \tag{1.6}$$

where we have taken into account that during the synchronization of clocks $C_0$ and $C_2$ the clock $C'$ has advanced with $V\dfrac{r}{c}$ in the positive direction of the common axes. We underline that (1.6) relates only physical quantities measured in the same reference frame K(XOY). Expressing (1.6) as a function of the readings of clocks $C'$ and $C_2$ when they meet, we obtain that the two readings are related by

$$t = \frac{t'}{\sqrt{1 - \dfrac{V^2}{c^2}}} = \gamma t' \tag{1.7}$$

using the shorthand notations $\gamma = \dfrac{1}{\sqrt{1-\beta^2}}; \beta = \dfrac{V}{c}$.

Figure 2.b shows the situation when the reflected synchronizing signal is received at the location of the origin O' which is when clock $C'_0$ reads $\dfrac{2d}{c} = 2t'$. At this very moment clock $C'_0$ is located in front of clock $C_3$ the first clock is reading $2t' = \dfrac{2d}{c}$ while the second clock is reading $2t$.

We underline that (1.7) holds only in the case when the initial position of the involved clocks is defined by a zero abscissa (in our case $x' = 0$).

Physicists call the reading of a clock *time coordinate*. They also use the concept of *time interval* that separates two events, defined as a difference between the readings of two clocks of the same inertial reference frame located where and when the events take place. Consider the situation when clock $C'_0$ meets clock $C_3$, as shown in Figure 3.

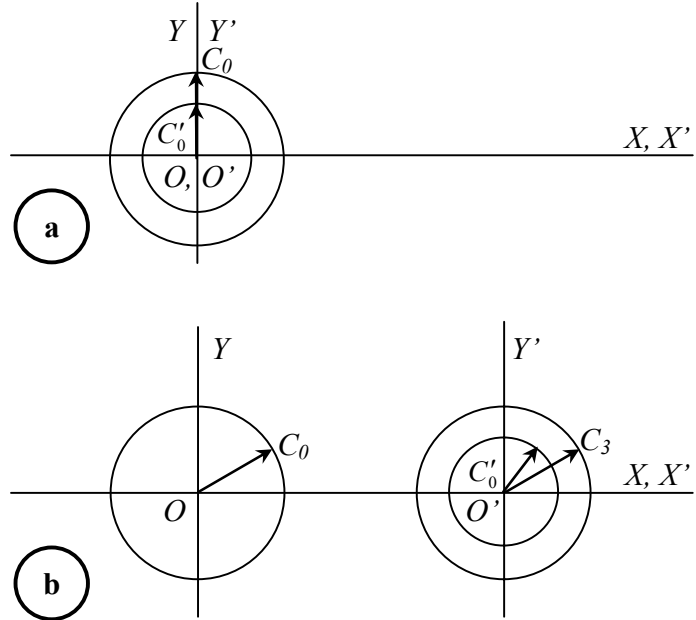

***Figure 3***. *Clock $C'_0$ meets clock $C_3$.*

Figure 3.a shows the situation when clock $C'_0$ is located in front of clock $C_0$ and they both read $t = t' = 0$. We consider that at that very moment the two clocks are synchronized by simple comparison, as they are set to read a zero time, a procedure that does not involve light signals.

Figure 3.b shows the situation when clock $C'_0$ arrives in front of a clock $C_3(x,0)$ of the K frame after a given time of motion; the first clock reads $t'$, while the second clock reads $t$. Therefore, we know that $x = Vt$. In accordance with the results obtained so far, we can consider that the two readings are related by (1.7), because $t$ and $t'$ represent clock readings i.e. time coordinates.

We have all the elements necessary to present an event as $E(x = r\cos\theta, r\sin\theta, t)$ where the space coordinates define the location of the point where it takes place, whereas the time coordinate $t$ equates the reading of the clock located at that point when the event takes place. We say that event $E$ and event $E'(x' = r'\cos\theta', y' = r'\sin\theta', t')$ represent the *same* event if the two events take place at the same point in space, where $t$ and $t'$ represent the readings of clocks of K and respectively K' located at the point where the events take place, both synchronized in their rest frames in accordance with Einstein's clock synchronization procedure.

The events involved in the experiment we have just described are $E_0(0,0,0)$ and $E'_0(0,0,0)$ associated with the fact that the clocks $C_0$ and $C'_0$ are located at the overlapped origins O and O' of the two reference frames and both read a zero time and also the events $E(x = Vt, 0, t)$ and $E'(0,0,t')$ associated with the fact that clock $C'_0$ arrives at the location of clock $C(x = Vt, 0)$ when the first reads $t'$ and the second reads $t$. Events $E_0$ and $E'_0$ are separated by a zero distance and by a zero time interval. Events $E$ and $E_0$ are separated by a time interval

$$t - 0 = \Delta t \tag{1.8}$$

and by a distance

$$x - 0 = \Delta x \tag{1.9}$$

obviously related by

$$\Delta x = V \Delta t. \tag{1.10}$$

The time interval $\Delta t$ is measured as a difference between the readings of two clocks $C$ and $C_0$ located at two different points in space. Events $E'_0$ and $E'$ are separated by a zero distance

$$\Delta x' = 0 \tag{1.11}$$

and by a time interval

$$\Delta t' = t' - 0 \tag{1.12}$$

which is measured as a difference between the readings of the same clock $C'_0$ in front of which both events take place. By definition, this represents a *proper time interval*. In accordance with (1.7), the time intervals are related by

$$\Delta t = \gamma \Delta t' \tag{1.13}$$

which relates the proper time interval $\Delta t'$ and the non-proper time interval $\Delta t$. Because $\Delta t > \Delta t'$ relativists say that a *time dilation effect* takes place.

So far, we have considered only events that take place on the overlapped axes ($y = y' = 0$). Returning to Figure 2a when clock $C'(0,d)$ reads $t' = \frac{d}{c}$, it is located in front of a clock $C_2(x = V\frac{r}{c}, d)$ which reads $t = \frac{r}{c}$. The problem is to find equations that relate the space-time coordinates of events $E'(x' = 0, y' = d, t' = \frac{d}{c})$ and $E(x = r\cos\theta, y = r\sin\theta, t = \frac{r}{c})$ associated with the fact that the clocks $C'$ and $C_2$ are located at the same point in space. We can define the location of clock $C'$ using polar coordinates so that $C'(r' = d, \theta' = 90^0)$. As we have seen, their time coordinates are related by

$$t = \gamma t'. \tag{1.14}$$

We also have
$$x = Vt = \gamma Vt' \tag{1.15}$$
$$y = y' = d \tag{1.16}$$
$$\cos\theta = \beta \tag{1.17}$$
$$\sin\theta = \gamma^{-1} \tag{1.18}$$
$$\tan\theta = \beta\gamma. \tag{1.19}$$

The invariance of distances measured perpendicular to the direction of relative motion requires that
$$r\sin\theta = d = y \tag{1.20}$$
i.e.
$$r = \gamma d. \tag{1.21}$$

The equations derived above are transformation equations that hold only in the case when one of the events takes place on the O'Y' axis of the reference frame K'. We can change the scenario we have followed so far, by considering the synchronization of clocks $C_0(0,0)$ and $C(0,d)$ of the K frame from the K' frame and by taking into account that K moves relative to K' in the negative direction of the common axes with speed $-V$. We obtain the *inverse transformation equations*
$$t' = \gamma t \tag{1.22}$$
$$x' = \gamma Vt \tag{1.23}$$
$$y = y' = d \tag{1.24}$$
$$\cos\theta' = -\beta \tag{1.25}$$
$$r' = \gamma d \tag{1.26}$$

which hold only when only **one** of the involved events takes place on the OY axis of the K frame.

The equations derived above enable observers of the two frames to measure the length of a moving rod. Consider a rod at rest in K and located along the common OX(O'X') axes with one of its ends **1**(0,0) located at the origin O. The length of the rod, measured by observers of its rest frame K, represents its *proper length* $L_0$. The position of the second end of the rod is defined by **2**($L_0$,0). Observers from K measure the speed of clock $C'_0$ that passes in front of end **1** when clock $C_0$ located there reads $t_1 = 0$ (Figure 4a) and passes in front of clock $C_2(L_0,0)$ located at end **2** when it reads $t_2 = t$. By definition the speed of clock $C'_0$ is
$$V = \frac{L_0}{t-0} = \frac{L_0}{\Delta t}. \tag{1.27}$$
where $\Delta t$ represents a non-proper time interval.

An observer $R'_0(0,0)$ located at the origin O' of his rest frame K' can use his wristwatch $C'_0(0,0)$ in order to measure the speed of the rod. The left end of the rod passes in front of him when his wristwatch reads $t'_1 = 0$ and the right end of the rod passes in front of him when the same clock reads $t'_2 = t'$. By definition the speed of the rod is

$$V = \frac{L}{t' - 0} = \frac{L}{\Delta t'} \qquad (1.28)$$

in which case $\Delta t'$ represents a proper time interval, whereas $L$ is a non-proper length. Combining (1.27) and (1.28) and taking into account (1.13) we obtain that the proper length $L_0$ and the *measured length* $L$ are related by

$$L = \gamma^{-1} L_0. \qquad (1.29)$$

It is important to underline that the time dilation formula relates a proper time interval and a non-proper time interval. It involves synchronized clocks only in the reference frame where the non-proper time interval is measured. The time dilation phenomenon loses most of its mystery once we recognize that it is basically the consequence of comparing successive readings in a given clock with readings in two different clocks.

**A.1.4 The Lorentz-Einstein transformations (LET)**

Due to the relative character of some physical quantities, we are not allowed to represent physical quantities measured in different inertial reference frames in relative motion on the same space diagram. With this in mind we no longer impose the restricting condition $x' = 0$ $(x = 0)$. Consider the events $E(x,0,t)$ and $E'(x',0,t')$ that take place somewhere on the overlapped axes OX(O'X'). Figure 4 shows the relative positions of the reference frames K and K' as detected from K when the synchronized clocks of that frame read $t$. A rod of proper length $L_0 = x' - 0$ at rest in K' is located along the common axes with one of its ends at O'. When representing it in 1.4a we should take into account that its length measured in K is

$$L = (x' - 0)\sqrt{1 - \frac{V^2}{c^2}} \qquad . \qquad (1.30)$$

Adding lengths measured in K we obtain

$$x = Vt + x'\sqrt{1 - \frac{V^2}{c^2}} \qquad (1.31)$$

from where we obtain

$$x' = \frac{x - Vt}{\sqrt{1 - \dfrac{V^2}{c^2}}} \quad . \tag{1.32}$$

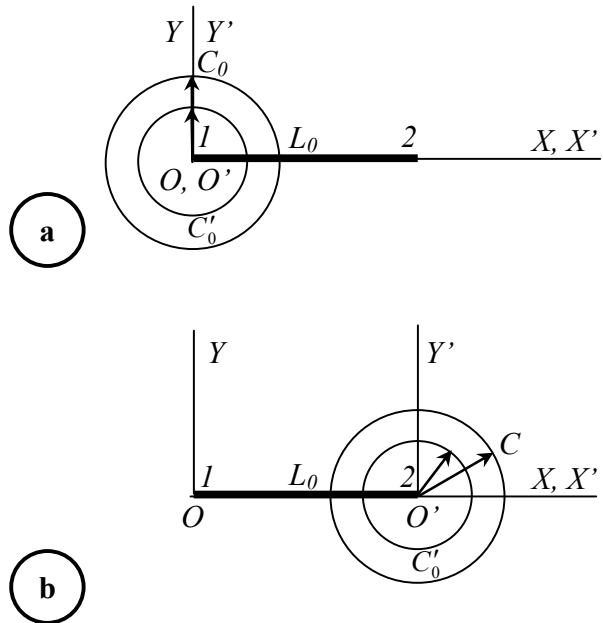

**Figure 4**. *The relative position of the reference frames K and K' as detected from K when the synchronized clocks of that frame read t.*

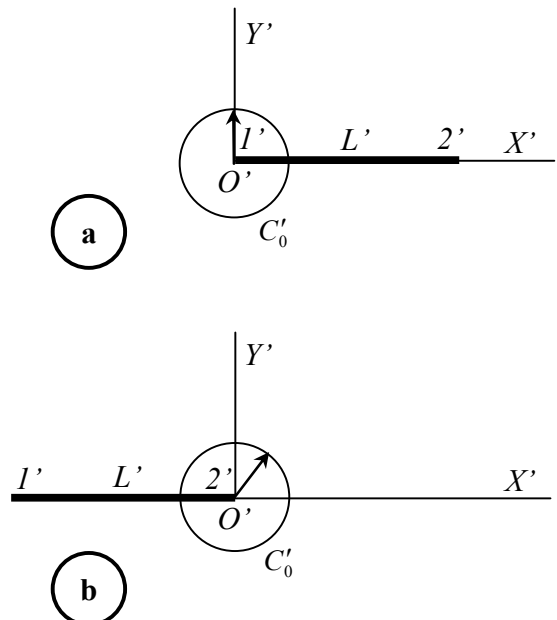

**Figure 5**. *The relative position of the reference frames K and K' as detected from K' when the synchronized clocks of that frame read t'.*

Figure 5 shows the relative position of the K and K' frames as detected from K' when the synchronized clocks of that frame read $t'$. A rod of proper length $L_0 = x - 0$ at rest in K has one of its ends at O. Its length measured in K' is

$$L = x\sqrt{1 - \frac{V^2}{c^2}}\ . \tag{1.33}$$

Adding length measured all in K' we obtain

$$x\sqrt{1 - \frac{V^2}{c^2}} = x' + Vt' \tag{1.34}$$

from which

$$x = \frac{x' + Vt'}{\sqrt{1 - \frac{V^2}{c^2}}}\ . \tag{1.35}$$

Combining (1.32) and (1.35) we obtain

$$t = \frac{t' + \frac{V}{c^2}t'}{\sqrt{1 - \frac{V^2}{c^2}}} \tag{1.36}$$

$$t' = \frac{t - \frac{V}{c^2}x}{\sqrt{1 - \frac{V^2}{c^2}}}\ . \tag{1.37}$$

We have derived the Lorentz-Einstein transformations for the space-time coordinates of events $E$ and $E'$ taking place at the same point of the overlapped axes when clocks $C(x,0)$ and $C'(x',0)$ located at that point read $t$ and respectively $t'$. By definition they represent the *same* event. From a practical point of view (1.35) and (1.36) enable observers from K' to express the space coordinates ($x,0$) of an event that takes place in K as a function of the space-time coordinates ($x',t'$) of the same event as detected from their rest frame K'. The practical value of (1.32) and (1.37) can be explained in the same way.

If the events defined as $E(x,y,t)$ in K and $E'(x',y',t')$ in K' take place at the same point located somewhere in the plane defined by the axes of the reference frames involved then we can add to the transformation equations derived above the equation

$$y = y' \tag{1.38}$$

which expresses the invariance of distances measured perpendicular to the direction of relative motion.

Consider a particle that starts to move at $t' = 0$ from the origin O' of the reference frame K' along a direction that makes an angle $\theta'$ with the positive direction of the common axes. The speed of this particle relative to K' is $\mathbf{u}(u_x, u_y)$. After a time of motion $t'$, the particle generates the event $E'(x' = u'_x t', = u't' \cos\theta', y' = u'_y t' \sin\theta', t')$. In accordance with the transformation equations derived above the space-time coordinates of the same event as detected from frame K are

$$x = \gamma(x' + Vt') = \gamma t'(u'_x + V) \tag{1.39}$$

or

$$x = \gamma(x' + V\frac{x'}{u'_x}) = \gamma x'(1 + \frac{V}{u'_x}) \tag{1.40}$$

and

$$t = \gamma(t' + \frac{V}{c^2}x') = \gamma t'(1 + \frac{Vu'_x}{c^2}) \tag{1.41}$$

or

$$t = \gamma(\frac{x'}{u'_x} + \frac{V}{c^2}x') = \gamma x'(1 + \frac{V}{c^2}u'_x). \tag{1.42}$$

Because the particle started to move at $t = t' = 0$ when the origins of the two frames were located at the same point in space, the components of their speeds relative to K are

$$u_x = \frac{x}{t} = \frac{x'}{t'}\frac{1 + \frac{V}{u'_x}}{1 + \frac{Vu'_x}{c^2}} = \frac{u'_x + V}{1 + \frac{Vu'_x}{c^2}} \tag{1.43}$$

$$u_y = \frac{y}{t} = \frac{y'}{t'}\frac{\gamma^{-1}}{1 + \frac{Vu'_x}{c^2}} = \frac{\gamma^{-1}u'_y}{1 + \frac{Vu'_x}{c^2}}. \tag{1.44}$$

Detected from K the particle moves along a direction $\theta$ relative to the positive direction of the common axes given by

$$\tan\theta = \frac{y}{x} = \frac{u_y}{u_x} = \frac{\gamma^{-1}\sin\theta'}{\cos\theta' + \frac{V}{u'}}. \tag{1.45}$$

We transform the magnitude of the speed as

$$u = \sqrt{u_x^2 + u_y^2} = u'\frac{\sqrt{(\cos\theta' + \frac{V}{u'})^2 + (1 - \frac{V^2}{c^2})\sin^2\theta'}}{1 + \frac{Vu'\cos\theta'}{c^2}}. \tag{1.46}$$

and the lengths of the position vector as

$$r = \sqrt{x^2 + y^2} = \gamma r'\sqrt{(\cos\theta' + \frac{V}{u'})^2 + (1 - \frac{V^2}{c^2})^2 \sin^2\theta'} \quad . \tag{1.47}$$

In many applications of special relativity we look for transformation equations that relate the space-time coordinates of events generated by light signals. We obtain them from the transformation equations derived above by taking into account that in this case, in accordance with the second postulate, $u = u' = c$. The result is

$$\tan\theta = \frac{\gamma^{-1}\sin\theta'}{\cos\theta' + \frac{V}{c}} \tag{1.48}$$

$$r = r'\gamma(1 + \frac{V}{c}\cos\theta') \tag{1.49}$$

$$t = t'\gamma(1 + \frac{V}{c}\cos\theta') \tag{1.50}$$

$$x = x'\sqrt{\frac{1 + \frac{V}{c}}{1 - \frac{V}{c}}} \tag{1.51}$$

The Lorentz-Einstein transformations derived above involve synchronized clocks in the inertial reference frames involved, K and respectively K'. Using different teaching strategies, authors derive the formulas that describe time dilation and length contraction starting with thought experiments, as we did, whereas others present them as consequences of the Lorentz-Einstein transformations[10].

We consider that the single well defined physical quantities with which special relativity operates are the *proper length* and the *proper time interval*. All the other concepts like *measured length* and *measured time interval* depend on the measurement procedure and so we could detect not only length contraction but also length dilation and not only time dilation but also time contraction.

### A.1.5 Conclusions
Underlining the part played by synchronized clocks of the Einstein type of observers we have derived the fundamental equations of relativistic kinematics in a compact way and from a little number of scenarios.